\documentclass[prd,nofootinbib,preprint,superscriptaddress,twocolumn,10pt]{revtex4}
\pdfoutput=1

\usepackage{amsmath,amssymb}
\usepackage{epsfig}
\usepackage{graphicx}
\usepackage[usenames,dvipsnames]{color}
\usepackage{subfigure}
\usepackage{slashed}
\usepackage{comment}
\usepackage[colorlinks,citecolor=blue]{hyperref}
\usepackage{color}
\newcommand{\lsim}{\mbox{\raisebox{-.6ex}{~$\stackrel{<}{\sim}$~}}}

\begin{document}
\title{Self-interacting dark matter and the GRB221009A event}

	\author{Debasish Borah}
	\affiliation{Department of Physics, Indian Institute of Technology Guwahati, Assam 781039, India}
	\author{Satyabrata Mahapatra}
	\affiliation{Department of Physics and Institute of Basic Science,
		Sungkyunkwan University, Suwon 16419, Korea}
	\author{Narendra Sahu}
	\affiliation{Department of Physics, Indian Institute of Technology Hyderabad, Kandi, Sangareddy 502285, Telangana, India}
	
	\author{Vicky Singh Thounaojam}
	\affiliation{Department of Physics, Indian Institute of Technology Hyderabad, Kandi, Sangareddy 502285, Telangana, India}
	\begin{abstract}
In this work, we explore the intriguing possibility of connecting self-interacting dark matter (SIDM) with the recently observed exceptionally bright and long-duration Gamma Ray Burst (GRB221009A). The proposed minimal scenario involves a light scalar mediator, simultaneously enabling dark matter (DM) self-interaction and explaining the observed very high energy (VHE) photons from GRB221009A reported by LHAASO's data. The scalar's mixing with the standard model (SM) Higgs boson allows for its production at the GRB site, which will then propagate escaping attenuation by the extra-galactic background light (EBL). These scalars, if highly boosted, have the potential to explain LHAASO's data. Moreover, the same mixing also facilitates DM-nucleon or DM-electron scatterings at terrestrial detectors, linking SIDM phenomenology to the GRB221009A events. This manuscript presents the parameter space meeting all constraints and offers an exciting opportunity to explore SIDM in future direct search experiments using insights from the GRB observation.

	\end{abstract}	
	\maketitle
	
\noindent
\section{Introduction}
The astrophysical and cosmological observations have always been one of the guiding principles while developing new physics models beyond the standard model (BSM), especially pertaining to dark matter (DM). For example, with the addition of DM, electromagnetism, and general relativity, correctly explain the observable spectrum of the cosmic microwave background (CMB) and the large-scale structure of the universe. Moreover, strong and weak interactions play non-trivial roles in the Big Bang Nucleosynthesis (BBN) of light nuclei in agreement with the observations. While there has been no discovery of BSM physics yet, there exist a few puzzling observations that may be hinting at some specific new physics scenarios. One such puzzle that has created a lot of intrigue among the particle physics community is the recently reported exceptionally bright and long-duration Gamma Ray Burst (GRB), GRB221009A \cite{SWIFT, FERMI-GBM, IPN, LHAASO, Burke-Gaffney,Fermi-GBM-Veres,AGILE/MCAL}.

On the 9th of October, 2022, this brightest GRB was detected by the Neil Gehrels Swift Observatory \cite{SWIFT}, Fermi satellite \cite{Fermi-LAT,Fermi-GBM-Veres, FERMI-GBM}, and various other detectors \cite{Burke-Gaffney, IPN,AGILE/MCAL}. In particular, Water Cherenkov Detector Array (WCDA) and Square kilometer array (KM2A) of Large High Altitude Air Shower Observatory (LHAASO \cite{LHAASO}) detected $\mathcal{O}(5000)$ events of photons with energies ranging from $0.5$~TeV to $18$~TeV within a time window of $2000\,s$ . However, these observations are puzzling since this has been detected at redshift $z = 0.15$ which corresponds to a comoving distance of $\sim600$ Mpc. Given this redshift of the source, the expected attenuation due to the extragalactic background light (EBL) is so severe that this detection has become very difficult to explain. The standard propagation model gives optical depths $\tau \sim$ 5(15) for the photons of energy, $E_\gamma \sim$ 10(18) TeV, and such a flux of VHE photons from extragalactic sources is strongly attenuated through the production of $e^{+} e^{-}$ pairs in the inter-galactic medium (IGM). This inconsistency motivated several attempts to solve the puzzle invoking physics beyond the Standard Model. Models involving Lorentz invariance violation \cite{Finke:2022swf}, axion-like particles \cite{Baktash, Galanti:2022pbg, Lin:2022ocj, Troitsky:2022xso, Nakagawa:2022wwm,Bernal:2023rdz}, sterile neutrinos \cite{Huang, Smirnov, Vedran, Guo}, light scalar \cite{Silk}, and external inverse Compton mechanism \cite{Zhang:2022lff, AlvesBatista:2022kpg, Gonzalez:2022opy} have appeared in the literature.

Another such long-standing puzzle stemming from the discrepancy between the astrophysical observations and $\Lambda$CDM simulations are the so-called small-scale structure problems namely core-cusp problem, too-big-to-fail and missing satellite problems~\cite{Spergel:1999mh,Tulin:2013teo, Kamada:2016euw, Creasey:2017qxc,Ren:2018jpt}. In light of these small-scale issues that the conventional cold dark matter (CDM) paradigm faces, the self-interacting dark matter (SIDM) offers an interesting viable alternative. In scenarios where DM possesses a light mediator, the necessary self-interaction, described in terms of the ratio of the cross-section to DM mass as $\sigma/m \sim 1$ cm$^2$/g can be naturally accomplished. Moreover, with the light mediator, one can achieve velocity-dependent DM self-interactions to address small-scale problems while maintaining consistency with typical CDM features at larger sizes.

\begin{figure}[h]
    \centering
    \includegraphics[scale=0.15]{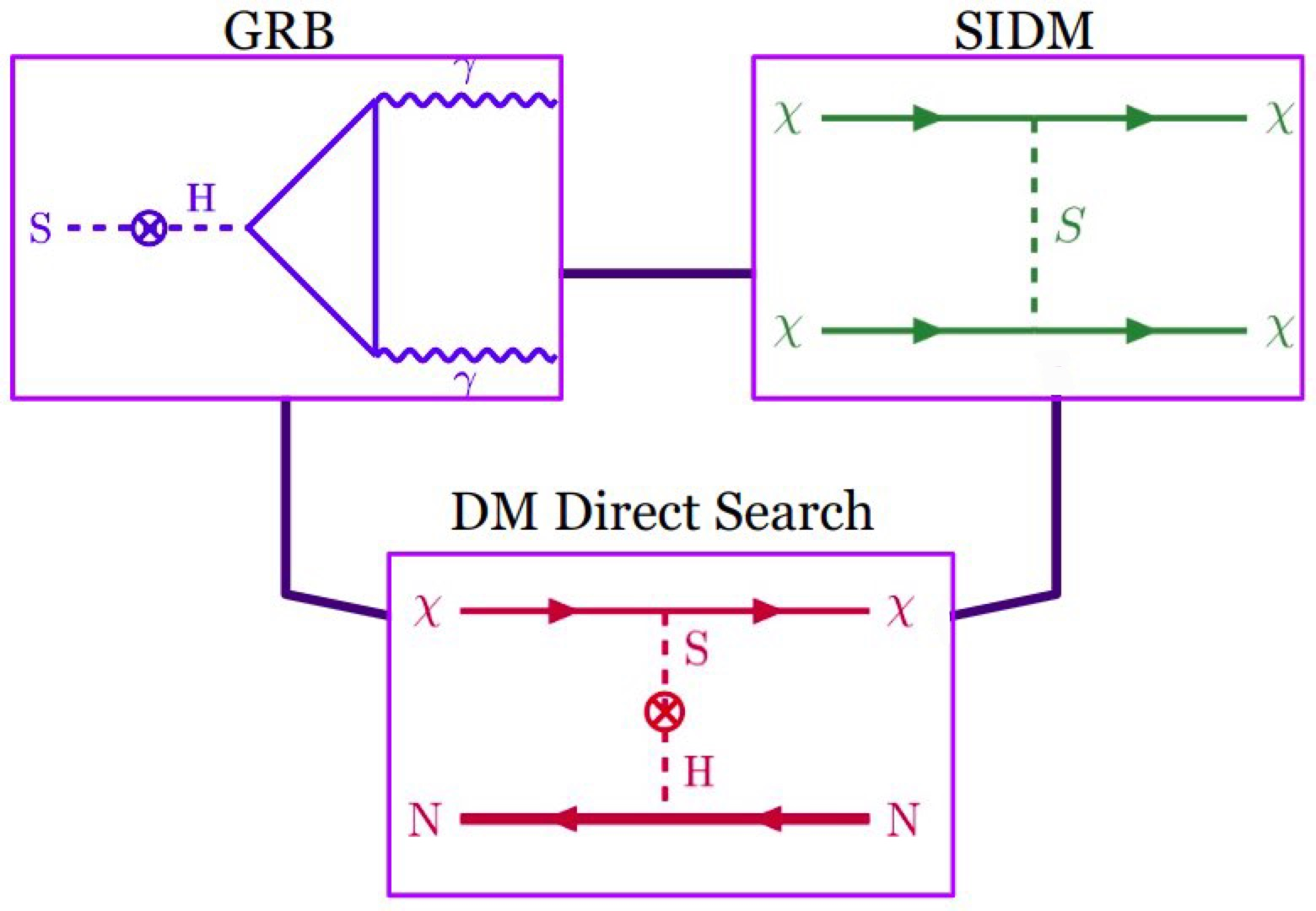}
    \caption{A schematic diagram to depict the key processes and their inter-connection in our scenario}
    \label{fig:schematic}
\end{figure}

In this article, we present an interesting scenario in which the self-interaction of DM can be probed both indirectly and in direct search experiments through the LHASSO's VHE photon events. We propose a minimal scenario where a light singlet scalar that mediates the DM self-scattering is also instrumental in explaining the VHE observation of GRB221009A evading the attenuation by the EBL. This is possible if the light scalar singlet $S$ mixes with the SM Higgs boson $h$ and thus the $S$ particles produced in the GRB are highly boosted. Once produced in the GRB, they can undergo radiative decay to two photons ({\it i.e.} $S \to \gamma \gamma$) while propagating to Earth. These resulting photons may thus be produced at a remote region without being nullified by the EBL. Interestingly, the mixing parameter between $S$ and $h$ which is crucial for the GRB221009A observation, also facilitates the DM-nucleon or DM-electron scatterings at the terrestrial detectors and hence establishes a unique connection between the DM phenomenology and GRB221009A observation. 

In Fig~\ref{fig:schematic}, we have shown the key processes for DM self-scattering, $S$ di-photon decay to explain the GRB events, and the DM-nucleon scattering at direct search experiments. The crucial feature to notice here is that not only the SIDM light mediator is responsible for explaining the GRB events, but the $S-H$ mixing angle that gives rise to the observed number of events can also be probed at the DM direct search experiments~\cite{GlobalArgonDarkMatter:2022xgs}.  

Here it is worth mentioning that, though recently there has been considerable interest in the regime of light DM typically in GeV or sub-GeV scale, mainly because such scenarios face weaker constraints from direct detection experiments \cite{LUX-ZEPLIN:2022qhg}, the DM coupling with light mediators to solve the small scale anomalies, results in significant DM annihilation rates, often leading to a lower-than-desired relic abundance in the low DM mass range below a few GeV. Despite various production mechanisms for SIDM proposed in the literature, obtaining the correct relic remains a formidable task. While a pure thermal relic is challenging to achieve, some studies have explored a hybrid approach involving both thermal and non-thermal contributions, which could potentially lead to the correct relic density for SIDM with the inclusion of new degrees of freedom making the model non-minimal\cite{Borah:2021yek, Borah:2021pet, Borah:2021rbx, Borah:2021qmi}. A new mechanism to achieve the thermal relic of SIDM by conversion from heavier partners during the freeze-out epoch is also recently proposed in~\cite{Borah:2022ask}.  In addition, the light scalar mediator also faces stringent cosmological constraints~\cite{Ibe:2021fed} because of its large lifetime required to explain the GRB221009A events by its di-photon decay. However, these two challenges can be mitigated if the mediator had a heavier mass in the early Universe~\cite{Elor:2021swj,Cohen:2008nb,Hashino:2021dvx}, which we discuss in subsequent sections.


\section{Minimal Setup}\label{model}
In this minimal setup, the SM particle content is augmented by a gauge singlet Dirac fermion $\chi$ and a singlet scalar $S$. $\chi$ is odd under an additional $Z_2$ symmetry while all other particles transform trivially under this $Z_2$. With its stability being guaranteed by the $Z_2$ symmetry, fermion $\chi$ becomes a viable dark matter candidate. The scalar $S$, because of its Yukawa coupling with $\chi$ mediates the dark matter self-interaction as shown in Fig.\ref{fig:schematic}. The relevant Lagrangian of this model can be written as:
\begin{equation}\label{eq1}
	\mathcal{L} \supset -M_\chi \overline{\chi} \chi-y_{S}\bar{\chi}\chi S + h.c. -V(H,S)
\end{equation}
and the relevant terms in the scalar potential is given by:
\begin{eqnarray*}
	V(H,S) &\supset& \mu_{S}^{2}S^{\dagger}S +\lambda_S (S^{\dagger}S)^{2} + \lambda_{SH}(H^{\dag}H)(S^{\dag}S)\nonumber\\ &+& \mu_{SH} S (H^\dag H) 
\end{eqnarray*}

Here it is worth mentioning that, the mixing that arises between the singlet scalar $S$ and the SM Higgs because of the couplings $\mu_{SH}$ and $\lambda_{SH}$, not only paves the way to detect DM at terrestrial laboratories but also explains the LHASSO's 18 TeV photon events.

In order for DM to possess a significant self-scattering cross section and to account for the GRB221009A events through its di-photon decay, it is necessary for $S$ to be light in the present day. 
We focus on DM mass in the range of sub-GeV to a few GeV which requires the $S$ to be in the MeV scale to achieve sufficient self-interaction that can resolve the small-scale anomalies of the Universe. Moreover, as discussed in section~\ref{section:grb}, to ensure that $S$ dominantly decays into two photons to explain the GRB events, the mass of $S$ is further constrained to be $M_S\leq 1$ MeV.

However, the presence of a light scalar with strong coupling to DM typically leads to a thermal underabundance of DM relic due to the strong annihilation of $\chi$ to $S$. Additionally, this scenario faces severe constraints as such a light scalar can be long-lived with a lifetime greater than the age of the Universe at the epoch of recombination and hence can distort the CMB anisotropy because of its decay to charged fermions or photons~\cite{Hambye:2019tjt}.
To address these challenges and salvage the scenario, it is preferable for the scalar $S$ to have a heavier mass during the early universe. Then, through a phase transition\footnote{We leave the details of such a phase transition and associated model building to an upcoming work.}, the initial mediator mass denoted as $M^i_S$ decreases to its current value $M^f_S = M_S \ll M^i_S$~\cite{Elor:2021swj, Cohen:2008nb}. For example, if mediator $S$ couples to another scalar $\eta$ driving a first-order phase transition (FOPT) with a coupling of the type $\mu \eta S^\dagger S$, then below the nucleation temperature of the FOPT, the physical mass of the mediator can change to $(M^f_S)^2 =(M^i_S)^2 -\mu v^2_\eta $ with $v_\eta$ being the vacuum expectation value (VEV) acquired by $\eta$. With suitable fine-tuning between the two terms $(M^i_S)^2, \mu v^2_\eta$, it is possible to get a final mediator mass that is the order of magnitude smaller than the initial mass. The discontinuous nature of FOPT can, therefore, abruptly reduce the mass of the mediator after the nucleation temperature. If this FOPT occurs well after establishing the correct relic abundance of DM, or DM freeze-out, then it does not affect the DM relic obtained at a higher temperature up to some dilution due to the release of latent heat from the FOPT. However, such dilution can be kept negligible depending upon the details of the FOPT model, which we leave for an upcoming work.

In this proposed scenario, the correct DM relic abundance can be achieved through the usual thermal freeze-out mechanism by suppressing the DM annihilation to $S$ through phase-space blocking, achieved by tuning the mediator mass $M^i_S$ according to $M_{DM}$. This also ensures that the lifetime of $S$ is consistent with the constraints from BBN which is otherwise of the order $\mathcal{O}(10^{13})$ s for a $1$ MeV scalar due to very small mixing with SM Higgs needed to explain the GRB events.


\section{GRB221009A}\label{section:grb}
In order to explain the LHASSO's GRB events, one first needs to know the production mechanism and energy profile of $S$ at the GRB site. Despite the environment, we still have the information for the most dominant mode for light scalar production, the nucleon-nucleon bremsstrahlung via pion exchange \cite{Krnjaic, Dev22, Dev20, Ishizuka, Arndt, Diener, Tu, Lee}. The relevant Lagrangian for this process can be written as: 
\begin{equation}\label{eq4}
	\mathcal{L}\supset -\sin\theta_{SH} \left[ A_\pi(\pi^0\pi^0 + \pi^+\pi^-) + y_{H}\overline{N}N + \frac{m_l}{v_{EW}}\bar{l}l \right]S
\end{equation}
and the process is
\begin{equation}\label{eq5}
	N + N \rightarrow N + N + S
\end{equation}
where N is either neutron(n) or proton (p) and $A_\pi$ is the effective coupling of $S$ to pions. Thus once produced, the scalar can decay into leptons or pions at tree-level or to photons at one loop level via mixing with the SM Higgs boson. The decay width of $S$ are given by \cite{Dev17}
\begin{equation}\label{eq6}
	\begin{split}
		\Gamma_{S\rightarrow \gamma \gamma} & =\frac{121}{9}\frac{\alpha^2 M_S^3 \sin^2\theta_{SH}}{512 \pi^3 v_{EW}^2}\\
		\Gamma_{S\rightarrow e^-e^+} & = \frac{M_S m_e^2 \sin^2\theta_{SH}}{8 \pi v_{EW}^2}\left(1-\frac{4m_e^2}{M_S^2} \right)^{3/2}
	\end{split}
\end{equation}

Now if the mass of S is at least twice the mass of electrons ($M_S\ge 2m_e$), then it will dominantly decay to $e^-e^+$ pairs. Thus, the ratio $\Gamma_{S\rightarrow e^-e^+}/\Gamma_{S\rightarrow \gamma \gamma}$ becomes very large and hence the di-photon decay is suppressed heavily. This will not be able to explain the LHAASO's data. Therefore, we restrict the mass of $S$ to be less than or equal to 1 MeV, and then, this $S$ scalar decays dominantly to di-photons. If $\Phi_S$ denotes the average flux of scalar S on Earth, the  corresponding gamma-ray flux from S decay is given by \cite{Silk}: 
\begin{equation}
	\Phi^S_\gamma=\frac{2 \Phi_S Br_\gamma}{\tau \lambda_S/d-1} \left( e^{-d/\lambda_S}-e^{-\tau} \right)
\end{equation}
where $d$ is the distance of the GRB source from Earth, $Br_\gamma$ is the branching fraction of $S\rightarrow\gamma\gamma$, and $\tau=d/ \lambda_\gamma$ is the optical depth of the photons with $\lambda_\gamma$ being the mean free path of the photon. The quantity $\lambda_S$ is given in terms of $S$ decay rate $\Gamma_S$ and the energy of the scalar $E_S$ in the Earth's rest frame
\begin{equation}
	\lambda_S=\frac{E_S}{M_S \Gamma_S}
\end{equation}
Given the information of this S-induced $\gamma$ flux, one can calculate the number of events as:
\begin{equation}
    N_\gamma =\Delta t \int_{0.5}^{18~{\rm TeV}}Area\times \Phi^S_\gamma(E_\gamma)dE_\gamma
\end{equation}
However, the un-attenuated gamma flux is obtained by extrapolating the flux measured by Fermi-LAT in the energy range from GeV to higher TeV as \cite{Baktash}
\begin{equation}
	\Phi_\gamma^0(E_\gamma)=\frac{2.1 \times10^{-6}}{\rm cm^2~s~TeV}\left( \frac{E_\gamma}{\rm TeV} \right)^{-1.87 \pm 0.04}
\end{equation}
\\
\begin{figure}[t]
	\includegraphics[scale=0.5]{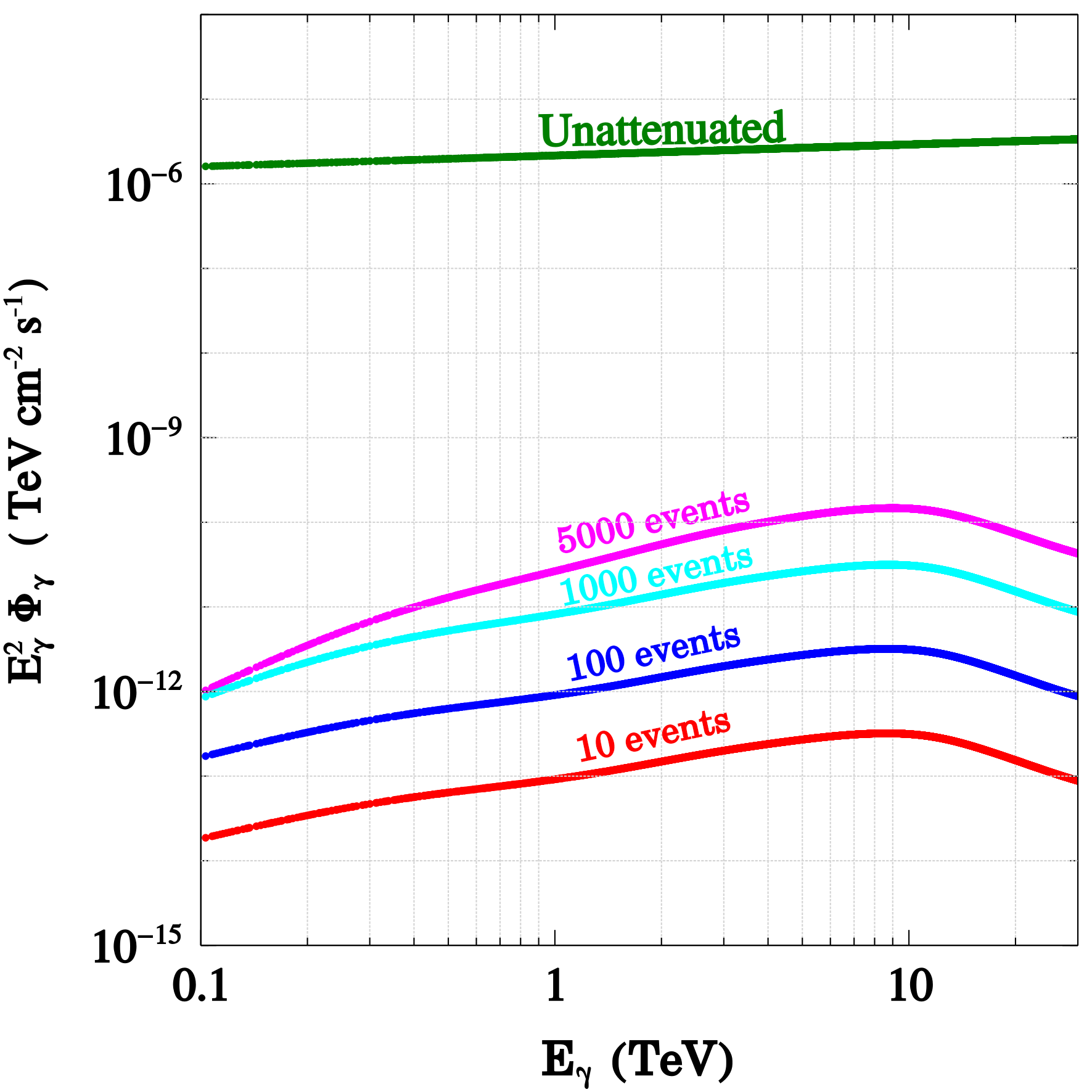}~
	 \caption{Flux of $S$-induced gamma rays for $N_\gamma =10,100,1000,\,5000$ assuming $E_S=2E_\gamma$ over a detector of area 1 ${\rm km}^2$ in a time window of $\Delta t=2000$ s. 
 }
	\label{fig:flux}
\end{figure}

In Fig~\ref{fig:flux}, we have shown the flux of $\gamma$-rays originated from the decay of $S$ as a function of the photon energy $E_\gamma$ that can give rise to 10, 100, 1000, and 5000 events assuming the scalar energy $E_S = 2~E_\gamma$. This flux has been calculated over a detector area of 1 ${\rm km}^2$ and in a time window of 2000 $s$ which is typical of LHASSO's KM2A detector. We have also shown the un-attenuated flux in the same figure by the red dotted line and it is clear that because of attenuation, the flux reduces by almost a factor of $\mathcal{O}(10^{-7})$ which is still sufficient to explain the TeV energy photons observed by LHASSO. 

\section{Self-Interacting DM}

It is well-known that a discrepancy exists between the results obtained from the simulation of collision-less cold dark matter and the astrophysical data at small scales. To address this mismatch, researchers have explored the possibility of relaxing the collision-less picture and considering self-scattering among the DM particles \cite{Kamada:2016euw, Creasey:2017qxc, Ren:2018jpt, Tulin:2013teo, Spergel:1999mh}.

By means of the $t$-channel process, the DM particle $\chi$ is capable of undergoing elastic self-scattering mediated by the scalar $S$. Notably, the typical cross-section for DM self-scattering in this scenario is significantly larger than that of typical thermal DM, and this effect arises naturally due to the presence of the light scalar $S$  which has a mass much smaller than typical weak-scale mediators. As discussed earlier in the preceding section, this scalar must have a mass restricted by the upper limit $M_{S} < 2 m_{e}$ in order to explain LHASSO's 18 TeV photon observation through the di-photon decay process. 

\begin{figure}[b]
	\centering
	\includegraphics[scale=0.5]{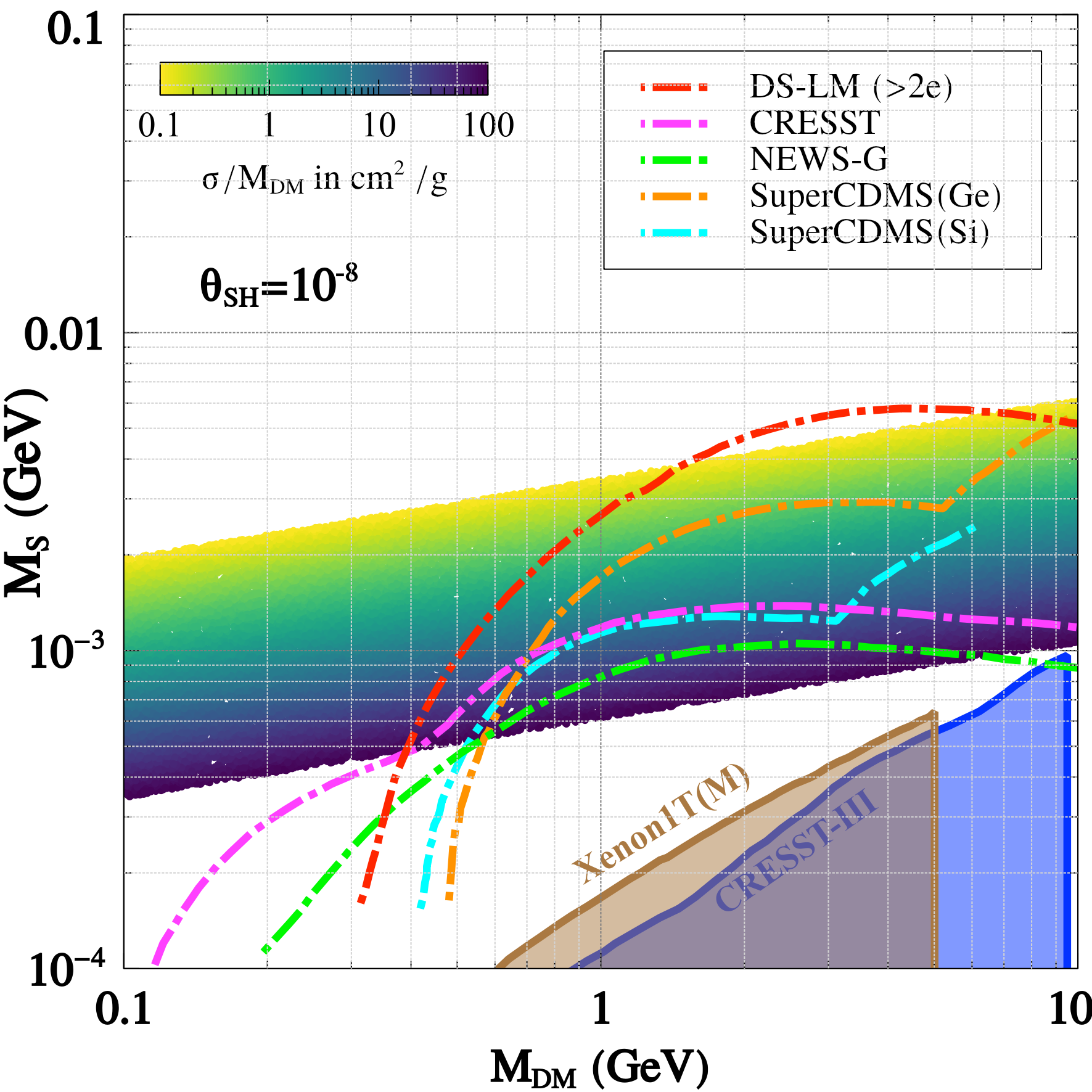}
	\caption{Constraints
from existing DM direct detection as well as the projected sensitivities of the future experiments in the plane of DM mass ($M_{\rm DM}$) versus mediator mass ($M_S$) for self-interaction.
}
\label{fig:ddcon}
\end{figure}

In the case of such a light mediator, the non-relativistic self-interaction of DM can be described by a Yukawa-type potential: $V(r)= \frac{y^2_S}{4\pi r}e^{-M_{S}r}$. By evaluating the quantum mechanical self-interaction cross sections and imposing constraints on $\sigma/M_{\rm DM}$ based on astrophysical observations at various scales, such as dwarfs, low surface brightness (LSB) galaxies, and clusters~\cite{Kaplinghat:2015aga, Kamada:2020buc}, we can determine the permissible parameter space of the model in the $M_{\rm DM}=M_{\chi}$ and mediator mass $M_S$ plane. This information is depicted in Fig.~\ref{fig:ddcon} by the color-coded band.
It is important to note that calculations need to extend beyond the perturbative limit to fully explore the parameter space, and based on the mass of the mediator, mass of DM, coupling strength, and relative velocity, three distinct regimes can be identified in this parameter space: the Born regime, the classical regime, and the resonant regime. Further details on the self-interaction of DM and the self-scattering cross-section can be found in~\cite{Tulin:2013teo,Tulin:2017ara,Tulin:2012wi,Khrapak:2003kjw}

\subsection{Production of SIDM}

As mentioned previously, in the simplest setup without introducing any new degrees of freedom into the theory, the thermal relic density of self-interacting dark matter (SIDM) particles ($\chi$) remains too low due to excessive annihilation into the light mediator, especially for DM in the GeV or sub-GeV mass range. However, by considering new particle species in thermal or hybrid scenarios, as discussed in~\cite{Borah:2021yek, Borah:2021pet, Borah:2021rbx, Borah:2021qmi, Borah:2022ask}, it becomes possible to achieve the correct relic density. 

In our study, we choose to work with the most minimal setup, where the dark sector consists solely of the dark matter $\chi$ and the mediator $S$. 
Note that as discussed in section~\ref{model}, the scalar $S$ goes through a phase transition in the early Universe. Prior to the phase transition the mass of $S$, $M^i_S\lsim M_{\rm DM}(=M_\chi)$, such that the DM annihilation rate to $S$ is phase-space suppressed. This reduces the $\chi\chi \to SS$ annihilation cross-section to give rise the correct relic density of $\chi$. After the phase transition, $S$ becomes light with $M_{S} \leq 1$ MeV that is required to describe the self-interaction of DM and the GRB events.    

\begin{figure}[h]
	\centering
	\includegraphics[scale=0.5]{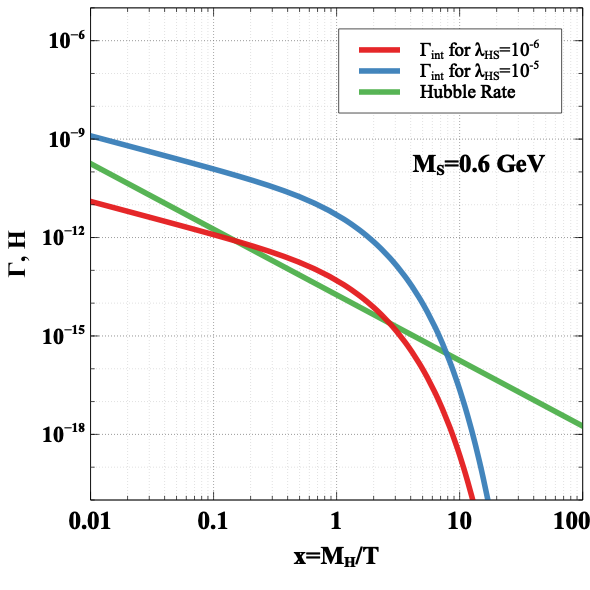}
	\caption{Interaction rate of $HH \leftrightarrow SS$ process in comparison to the Hubble expansion rate for two different values of $\lambda_{SH}$. }
	\label{fig:thermal}
\end{figure}

In our scenario, the singlet scalar $S$ is in thermal equilibrium with the SM bath through its scalar portal interactions. This also brings DM $\chi$ to thermal equilibrium because of its   significant coupling with $S$.
The $S-H$ mixing angle is given by: 
\begin{equation}
    \tan(2\theta_{SH})  =  \frac{(uv \lambda_{SH} + \sqrt{2}\, v\,\mu_{SH})}{M_H^2 - M_S^2}
\end{equation}

We maintain $\theta_{SH} = \mathcal{O}(10^{-8})$ by choosing $\lambda_{SH}$ and $\mu_{SH}$ appropriately. In particular, in Fig~\ref{fig:thermal}, we show that  for $\lambda_{SH}\sim \mathcal{O}(10^{-5})$, $HH \leftrightarrow SS$ can be in thermal equilibrium while still being consistent with the perturbativity, unitarity and co-positivity constraints. By setting $u=1$ GeV, we choose two benchmark points for 
$\{\lambda_{SH},\mu_{SH}\}$ {\it i.e.}
$\{10^{-5},-6.6 \times 10^{-6}\}$ and $\{10^{-6},-2.6 \times 10^{-7}\}$ which corresponds to $\theta_{SH}=10^{-8}$ and showcase the interaction rate for the process $HH \leftrightarrow SS$ against the Hubble expansion rate in Fig.~\ref{fig:thermal}. Clearly this justifies the assumption that $S$ and $DM$ are in thermal equilibrium in the early Universe.

\begin{figure}[h]
	\centering
	\includegraphics[scale=0.5]{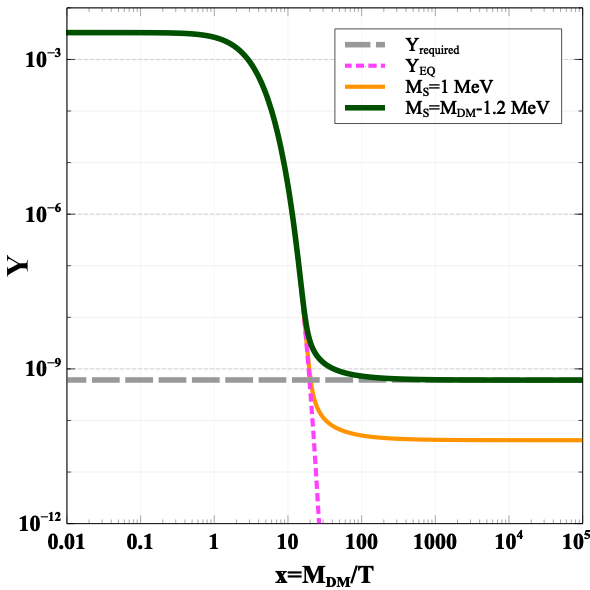}
	\caption{Evolution of co-moving number density of DM of mass $M_{DM}=0.7\, GeV$.}
	\label{fig:freezeout}
\end{figure}
 
In Fig.~\ref{fig:freezeout}, we show the evolution of the co-moving number density of DM  for $M_{DM}=0.7$ GeV by solving the relevant Boltzmann equation:
\begin{equation}\label{eq11}
    \frac{dY_{DM}}{dx} = -\frac{s(M_{DM}) \langle \sigma v \rangle_{total}}{x^2 H(M_{DM})}\left(Y_{DM}^2-(Y^{eq}_{DM})^2\right)
\end{equation}
 where $x=M_{DM}/T$, $H$ is the Hubble parameter, and $s$ is the entropy density. 
Here $\langle \sigma v \rangle_{total}=\langle \sigma v \rangle_{\chi \chi \to S S}+\langle \sigma v \rangle_{\chi \chi \to SM SM}$. However, due to the small mixing angle $\theta_{SH}$, DM annihilation into SM final state processes are less significant compared to the $\chi \chi \to S S$ process, which is governed by the same coupling $y_S$ that is responsible for DM self-interaction.
The thermally averaged cross-section for $\chi$ annihilation to $S$, which is the dominant number changing process for $\chi$, is given by:
\begin{equation}
    \langle\sigma v\rangle = \frac{3}{4} \frac{y^2_S}{16 \pi M^2_\chi} v^2 \left(1-\frac{(M^{i}_S)^2}{M^2_\chi}\right)^{1/2}
    \label{eq:dmann}
\end{equation}
 Here it is worth noticing that for a particular value of $M_\chi$ and $y_S$, which are relevant for self-interaction at late times, we can achieve the correct relic abundance of $\chi$ by appropriately choosing the value of $M^i_S$.
In particular, we have chosen $M_{\chi}-M^{i}_S=1.2$ MeV for $M_{\chi}=0.7$ GeV and $y_S=0.03$. The corresponding comoving number density of $\chi$ is shown by the green line in Fig~\ref{fig:freezeout}. We see that the correct abundance of DM can be achieved with the above mentioned parameters.
 
However, had the $M_S$ value been chosen to be $M^i_{S}=M^{f}_{S}=1$ MeV, with the same choice of $M_\chi=0.7$ GeV and $y_S=0.03$, then the resulting DM abundance would have been two orders of magnitude smaller than the observed relic due to efficient annihilation of $\chi$ into $S$. This is depicted by the orange line in Fig~\ref{fig:freezeout}.

\subsection{Direct Detection}
As mentioned earlier, the most intriguing link between the observation of GRB221009A and SIDM in our scenario emerges when we explore the possibilities of direct detection of SIDM in terrestrial laboratories. The spin-independent elastic scattering of DM-nucleon becomes possible due to $S-H$ mixing, which is significantly influenced by $M_S$ and $\theta_{SH}$ for a fixed DM mass and $y_S$ coupling. Interestingly, this same $S-H$ mixing plays a crucial role not only in the production of $S$ particles at the GRB site but also in their subsequent di-photon decay, which offers a plausible explanation for the observation of 18 TeV photon by the LHASSO experiment.

This fascinating connection allows us to explore the SIDM parameter space in direct search experiments with the added advantage of using the information on $\theta_{SH}$ obtained from the GRB events. Normally, $\theta_{SH}$ is considered a free parameter in the theory, but the GRB observation provides valuable insights, making it a more constrained and informative aspect of this study as demonstrated in Fig.~\ref{fig:schematic} too.

 The spin-independent (SI) scattering cross-section of DM per nucleon is given by:
 \begin{equation}\label{eq7}
 	\sigma_{\rm SI}=\frac{\mu_r^2}{4\pi A^2}[Zf_p + (A-Z)f_n]^2
 \end{equation}

 where $\mu_r$ is the reduced mass of the DM-nucleon system. $A$ and $Z$ are the mass number and the atomic number of the target nucleus respectively. The interaction strengths $f_p$ and $f_n$ of proton and neutron with DM are given as:
 \begin{equation}\label{eq8}
 	f_{p,n}=\sum_{q=u,d,s} f_{T_q}^{p,n}\alpha_q \frac{m_{p,n}}{mq} + \sum_{q=c,t,b} f_{TG}^{p,n}\alpha_q \frac{m_{p,n}}{mq}
 \end{equation}
 with $\alpha_q$ defined by 
 \begin{equation}
 	\alpha_q=y_s \sin\theta_{SH}\frac{m_q}{v_{EW}}\left[ \frac{1}{M_S^2}-\frac{1}{M_H^2}\right]
 \end{equation}
 \\
 Here it is worth mentioning that for the explanation of LHASSO's VHE photon events, the mixing parameter $\theta_{SH}$ which is consistent with all the relevant constraints is of the order $\mathcal{O}(10^{-8})$. Thus we take the benchmark values of $\theta_{SH}=10^{-8}$ and $y_{S}=0.03$ and showcase the constraints from direct search experiments against the parameter space giving rise to the required self-interaction cross-section for the dwarf galaxies ($v \sim 10$ km/s) on the plane of $M_S ~{\rm vs}~M_{DM}$ in the Fig.~\ref{fig:ddcon}.
 The shaded regions in the plot represent the existing constraints from CRESST-III~\cite{CRESST:2019jnq} and Xenon1T(M)\cite{Billard:2021uyg, Ibe:2017yqa}. These regions do not exclude the self-interaction parameter space for $\theta_{SH}=10^{-8}$. However, it is noteworthy that the sensitivity projections (at 90\% C.L.) for spin-independent DM-nucleon scattering of CRESST\cite{CRESST:2019jnq}, NEWS-G~\cite{NEWS-G:2017pxg}, SuperCDMS~\cite{SuperCDMS:2016wui}, and DS-LM~\cite{GlobalArgonDarkMatter:2022ppc} experiments have the potential to significantly probe the SIDM parameter space corresponding to the $S-H$ mixing angle inferred from LHASSO's observation of VHE photons.

\section{Summary and Conclusion}
In this \textcolor{red}{article}, we unveil a fascinating and innovative link between SIDM and the recently observed VHE photon events associated with GRB221009A, as reported by LHASSO. Our study centers on the most minimal configuration, where the dark sector comprises one fermionic DM particle, denoted as $\chi$, and a light scalar mediator, referred to as $S$. This setup is designed to introduce velocity-dependent self-interaction among DM, addressing small-scale issues.

Interestingly, the presence of the light scalar mediator plays a pivotal role in explaining the GRB events. During the occurrence of GRB, this light scalar is produced at the GRB site through its mixing with the SM Higgs. Subsequently, it undergoes significant boosting, remaining unattenuated by the EBL. Upon reaching Earth, the light scalar decays remotely, yielding two highly energetic photons via radiative processes. These photons then manifest as VHE signals, detectable by terrestrial detectors.


\begin{figure}[h]
\includegraphics[scale=0.5]{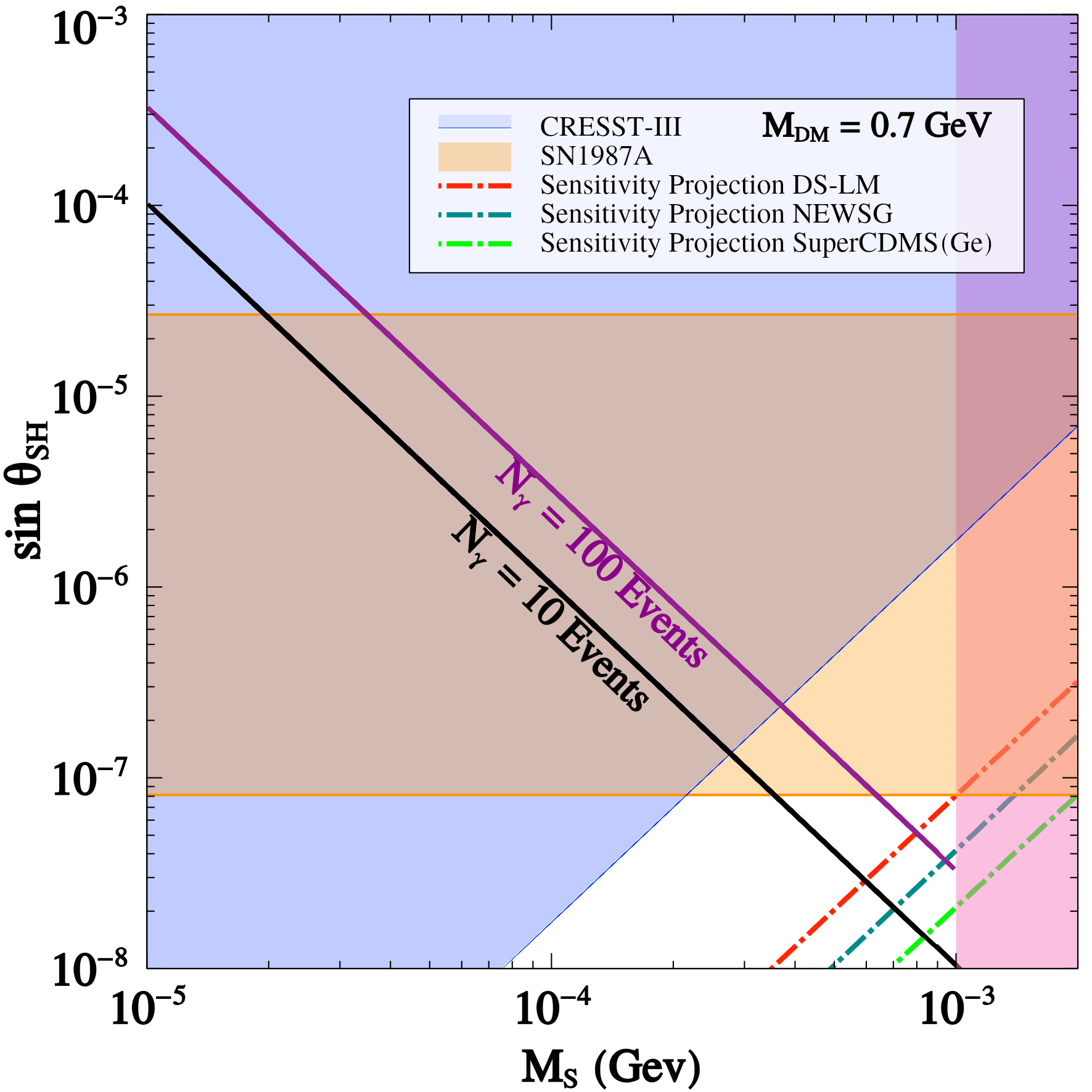}
\caption{The final parameter space to explain the LHAASO's 18 TeV photon observation.}
\label{fig:ps}
\end{figure}

In Figure~\ref{fig:ps}, we present the final parameter space that explains the GRB221009A observation, considering all relevant constraints. The solid black and purple contours represent the parameter space capable of yielding an expected number of VHE photon events, $N_\gamma=10$ and $N_\gamma=100$ respectively. The orange shaded band shows the constraints from SN1987A, which arise because abundant production of light mediators $S$ leaving the source conflicts with the observed neutrino luminosity~\cite{Balaji22}. Notably, the most stringent constraint originates from DM direct search. The blue shaded region denotes the constraints from CRESST-III~\cite{CRESST:2019jnq} on SI DM-nucleon scattering, ruling out larger values of $\sin\theta_{SH}$ and smaller $M_{S}$ below $100$ keV. Additionally, we illustrate the projected sensitivities of future direct search experiments DS-LM~\cite{GlobalArgonDarkMatter:2022ppc}, NEWS-G~\cite{NEWS-G:2017pxg}, and SuperCDMS~\cite{SuperCDMS:2016wui} with dot-dashed contours, indicating their potential to explore the parameter space further.

We emphasized that the correct relic density of SIDM can be achieved through the thermal freeze-out mechanism by tuning the mediator mass to a heavier value in the early universe. This adjustment addresses the issue of under-abundance due to excessive annihilation to $S$. Subsequently, the mass of $S$ can decrease to its present value after a phase transition that occurs well after the establishment of DM relic density and $S$ decay to SM fermions via $S-H$ mixing which reduces the $S$ number density $Y_{S} \to 0$. Consequently, our scenario remains consistent with BBN and CMB constraints on light mediator decay~\cite{Hambye:2019tjt}. Importantly, our case benefits from $p$-wave suppressed DM annihilation to light mediators (as mentioned in Eq.\ref{eq:dmann}), making it safe from constraints on DM annihilation during the recombination epoch\cite{Hambye:2019tjt}.

By satisfying all these constraints, the remarkable connection between the GRB events and SIDM enables us to explore the SIDM parameter space in future DM direct search experiments, utilizing the information of $\theta_{SH}$ inferred from the GRB observation. This unique connection enhances the predictiveness of the entire scenario.



\acknowledgments
The work of NS is supported by the Department of Atomic Energy-Board of Research in Nuclear Sciences, Government of India (Ref. Number: 58/14/15/2021- BRNS/37220). SM acknowledges the support from the National Research Foundation of Korea grant 2022R1A2C1005050. The work of DB is supported by the Science and Engineering Research Board (SERB), Government of India grant MTR/2022/000575. 
 
 \section*{}
 \appendix

 \section{Low energy self-interaction cross sections}\label{appendixB}
 The non-relativistic DM selg-scattering can be well understood in terms of the attractive Yukawa potential
 \begin{equation}
     V(r)=\frac{y_s^2}{4\pi r}e^{-M_Sr}
 \end{equation}
 To capture the relevant physics of forward scattering, the transfer cross-section is defined as
 \begin{equation*}
     \sigma_T=\int d\Omega (1-cos\theta)\frac{d\sigma}{d\Omega}
 \end{equation*}
 In the Born limit, $y_s^2M_{DM}/(4\pi s)\ll 1$,
     \begin{equation*}
     \sigma^{\rm Born}_T=\frac{y_s^4}{2\pi M_DM^2 v^4} \left[ log\left( 1+\frac{M_{DM}^2 v^2}{M_S^2} \right)-\frac{M_{DM}^2 v^2}{M_S^2+M_{DM}^2 v^2} \right]
 \end{equation*}
 Outside the Born limit, where   $y_s^2M_{DM}/4\pi s \geq 1$, there can be two different regions: classical regime and resonance regime. In the classical regime (${M_{DM}v}/{M_S}\geq 1$), solutions for an attractive potential is given by \cite{Tulin:2012wi, Tulin:2013teo, PhysRevLett.90.225002}
     \begin{equation*}
   \sigma^{\rm class.}_T=\begin{cases}
     \frac{4\pi}{M_S^2}\beta^2 ln(1+\beta^{-1}) & \textbf{$\beta > 1$}\\
     \frac{8\pi}{M_S^2}\left[ \frac{\beta^2}{1+1.5\beta^{1.65}} \right] & \textbf{$10^{-1} < \beta \leq 10^3$} \\
     \frac{\pi}{M_S^2}\left[ ln\beta +1-1/2 ln^{-1}\beta)\right]^2 & \textbf{$\beta \geq 10^3$}
   \end{cases}
 \end{equation*}
 where $\beta=\frac{2y_s^2M_{DM}}{4\pi M_{S}v^2}$. 

 Finally in the resonance region (${M_{DM}v}/{M_S}\leq 1$), no analytical formula for $\sigma_T$ is available. So approximating the Yukawa potential by Hulthen potential $\left(V(r)=\pm \frac{y^2_s}{4\pi}\frac{\delta e^{-\delta r}}{1-e^{-\delta r}}\right)$, the transfer cross-section is obtained to be: 
     \begin{equation*}
         \sigma_T^{Hulthen}=\frac{16\pi sin^2\delta_0}{M_{DM}^2v^2}
     \end{equation*}
 where $l=0$ phase shift $\delta_0$ is given by:
 $$ \delta_0=Arg \left[ \frac{i\Gamma(iM_{DM}v/kM_S)}{\Gamma(\lambda_+)\Gamma(\lambda_-)} \right]$$
 {\rm with}
 $$\lambda_{\pm}=1+\frac{iM_{DM}v}{2kM_S}\pm \sqrt{\frac{y_s^2 M_{DM}}{4\pi kM_S}-\frac{M_{DM}^2v}{4k^2M_S}}$$
 and k $\approx$ 1.6 is a dimensionless number. 


	\twocolumngrid

\begin{thebibliography}{63}
	\expandafter\ifx\csname natexlab\endcsname\relax\def\natexlab#1{#1}\fi
	\expandafter\ifx\csname bibnamefont\endcsname\relax
	\def\bibnamefont#1{#1}\fi
	\expandafter\ifx\csname bibfnamefont\endcsname\relax
	\def\bibfnamefont#1{#1}\fi
	\expandafter\ifx\csname citenamefont\endcsname\relax
	\def\citenamefont#1{#1}\fi
	\expandafter\ifx\csname url\endcsname\relax
	\def\url#1{\texttt{#1}}\fi
	\expandafter\ifx\csname urlprefix\endcsname\relax\def\urlprefix{URL }\fi
	\providecommand{\bibinfo}[2]{#2}
	\providecommand{\eprint}[2][]{\url{#2}}
	
	\bibitem[{\citenamefont{Kennea and (PSU)}(2022)}]{SWIFT}
	\bibinfo{author}{\bibfnamefont{J.~A.} \bibnamefont{Kennea}} \bibnamefont{and}
	\bibinfo{author}{\bibfnamefont{M.~W.} \bibnamefont{(PSU)}}
	(\bibinfo{year}{2022}),
	\urlprefix\url{https://gcn.gsfc.nasa.gov/gcn3/32635.gcn3}.
	
	\bibitem[{\citenamefont{Team}(2022{\natexlab{a}})}]{FERMI-GBM}
	\bibinfo{author}{\bibfnamefont{F.~G.} \bibnamefont{Team}}
	(\bibinfo{year}{2022}{\natexlab{a}}),
	\urlprefix\url{https://gcn.nasa.gov/circulars/32642.txt}.
	
	\bibitem[{\citenamefont{Konus-Wind~team}(2022)}]{IPN}
	\bibinfo{author}{\bibfnamefont{I.~S.-A. G. t. S.-B.~t.}
		\bibnamefont{Konus-Wind~team}, \bibfnamefont{Fermi GBM~team}}
	(\bibinfo{year}{2022}),
	\urlprefix\url{https://gcn.gsfc.nasa.gov/gcn3/32641.gcn3}.
	
	\bibitem[{\citenamefont{Yong~Huang and report on behalf of~the
			LHAASO~experiment}(2022)}]{LHAASO}
	\bibinfo{author}{\bibfnamefont{S.~C. M. Z. C. L. Z.~Y.}
		\bibnamefont{Yong~Huang}, \bibfnamefont{Shicong~Hu}} \bibnamefont{and}
	\bibinfo{author}{\bibfnamefont{Z.~C.} \bibnamefont{report on behalf of~the
			LHAASO~experiment}} (\bibinfo{year}{2022}),
	\urlprefix\url{https://gcn.gsfc.nasa.gov/gcn3/32677.gcn3}.
	
	\bibitem[{\citenamefont{optical observations}(2022)}]{Burke-Gaffney}
	\bibinfo{author}{\bibfnamefont{B.-G.~O.} \bibnamefont{optical observations}}
	(\bibinfo{year}{2022}),
	\urlprefix\url{https://gcn.gsfc.nasa.gov/gcn3/32664.gcn3}.
	
	\bibitem[{\citenamefont{Team}(2022{\natexlab{b}})}]{Fermi-GBM-Veres}
	\bibinfo{author}{\bibfnamefont{T.~F.~G.} \bibnamefont{Team}}
	(\bibinfo{year}{2022}{\natexlab{b}}),
	\urlprefix\url{https://gcn.nasa.gov/circulars/32636.txt}.
	
	\bibitem[{\citenamefont{Team}(2022{\natexlab{c}})}]{AGILE/MCAL}
	\bibinfo{author}{\bibfnamefont{T.~A.} \bibnamefont{Team}}
	(\bibinfo{year}{2022}{\natexlab{c}}),
	\urlprefix\url{https://gcn.nasa.gov/circulars/32650.txt}.
	
	\bibitem[{\citenamefont{(Politecnico and INFN~Bari)}(2022)}]{Fermi-LAT}
	\bibinfo{author}{\bibfnamefont{E.~B.} \bibnamefont{(Politecnico}}
	\bibnamefont{and} \bibinfo{author}{\bibfnamefont{M.~K.~N.}
		\bibnamefont{INFN~Bari)}, \bibfnamefont{N.~Omodei (Stanford~Univ.)}}
	(\bibinfo{year}{2022}),
	\urlprefix\url{https://gcn.nasa.gov/circulars/32637.txt}.
	
	\bibitem[{\citenamefont{Finke and Razzaque}(2023)}]{Finke:2022swf}
	\bibinfo{author}{\bibfnamefont{J.~D.} \bibnamefont{Finke}} \bibnamefont{and}
	\bibinfo{author}{\bibfnamefont{S.}~\bibnamefont{Razzaque}},
	\bibinfo{journal}{Astrophys. J. Lett.} \textbf{\bibinfo{volume}{942}},
	\bibinfo{pages}{L21} (\bibinfo{year}{2023}), \eprint{2210.11261}.
	
	\bibitem[{\citenamefont{Baktash et~al.}(2022)\citenamefont{Baktash, Horns, and
			Meyer}}]{Baktash}
	\bibinfo{author}{\bibfnamefont{A.}~\bibnamefont{Baktash}},
	\bibinfo{author}{\bibfnamefont{D.}~\bibnamefont{Horns}}, \bibnamefont{and}
	\bibinfo{author}{\bibfnamefont{M.}~\bibnamefont{Meyer}}
	(\bibinfo{year}{2022}), \eprint{2210.07172}.
	
	\bibitem[{\citenamefont{Galanti et~al.}(2022)\citenamefont{Galanti, Nava,
			Roncadelli, and Tavecchio}}]{Galanti:2022pbg}
	\bibinfo{author}{\bibfnamefont{G.}~\bibnamefont{Galanti}},
	\bibinfo{author}{\bibfnamefont{L.}~\bibnamefont{Nava}},
	\bibinfo{author}{\bibfnamefont{M.}~\bibnamefont{Roncadelli}},
	\bibnamefont{and} \bibinfo{author}{\bibfnamefont{F.}~\bibnamefont{Tavecchio}}
	(\bibinfo{year}{2022}), \eprint{2210.05659}.
	
	\bibitem[{\citenamefont{Lin and Yanagida}(2022)}]{Lin:2022ocj}
	\bibinfo{author}{\bibfnamefont{W.}~\bibnamefont{Lin}} \bibnamefont{and}
	\bibinfo{author}{\bibfnamefont{T.~T.} \bibnamefont{Yanagida}}
	(\bibinfo{year}{2022}), \eprint{2210.08841}.
	
	\bibitem[{\citenamefont{Troitsky}(2022)}]{Troitsky:2022xso}
	\bibinfo{author}{\bibfnamefont{S.~V.} \bibnamefont{Troitsky}},
	\bibinfo{journal}{Pisma Zh. Eksp. Teor. Fiz.} \textbf{\bibinfo{volume}{116}},
	\bibinfo{pages}{745} (\bibinfo{year}{2022}), \eprint{2210.09250}.
	
	\bibitem[{\citenamefont{Nakagawa et~al.}(2023)\citenamefont{Nakagawa,
			Takahashi, Yamada, and Yin}}]{Nakagawa:2022wwm}
	\bibinfo{author}{\bibfnamefont{S.}~\bibnamefont{Nakagawa}},
	\bibinfo{author}{\bibfnamefont{F.}~\bibnamefont{Takahashi}},
	\bibinfo{author}{\bibfnamefont{M.}~\bibnamefont{Yamada}}, \bibnamefont{and}
	\bibinfo{author}{\bibfnamefont{W.}~\bibnamefont{Yin}},
	\bibinfo{journal}{Phys. Lett. B} \textbf{\bibinfo{volume}{839}},
	\bibinfo{pages}{137824} (\bibinfo{year}{2023}), \eprint{2210.10022}.
	
	\bibitem[{\citenamefont{Bernal et~al.}(2023)\citenamefont{Bernal, Farzan, and
			Smirnov}}]{Bernal:2023rdz}
	\bibinfo{author}{\bibfnamefont{N.}~\bibnamefont{Bernal}},
	\bibinfo{author}{\bibfnamefont{Y.}~\bibnamefont{Farzan}}, \bibnamefont{and}
	\bibinfo{author}{\bibfnamefont{A.~Y.} \bibnamefont{Smirnov}}
	(\bibinfo{year}{2023}), \eprint{2307.10382}.
	
	\bibitem[{\citenamefont{Huang et~al.}(2023)\citenamefont{Huang, Wang, Yu, and
			Zhou}}]{Huang}
	\bibinfo{author}{\bibfnamefont{J.}~\bibnamefont{Huang}},
	\bibinfo{author}{\bibfnamefont{Y.}~\bibnamefont{Wang}},
	\bibinfo{author}{\bibfnamefont{B.}~\bibnamefont{Yu}}, \bibnamefont{and}
	\bibinfo{author}{\bibfnamefont{S.}~\bibnamefont{Zhou}},
	\bibinfo{journal}{JCAP} \textbf{\bibinfo{volume}{04}}, \bibinfo{pages}{056}
	(\bibinfo{year}{2023}), \eprint{2212.03477}.
	
	\bibitem[{\citenamefont{Smirnov and Trautner}(2022)}]{Smirnov}
	\bibinfo{author}{\bibfnamefont{A.~Y.} \bibnamefont{Smirnov}} \bibnamefont{and}
	\bibinfo{author}{\bibfnamefont{A.}~\bibnamefont{Trautner}}
	(\bibinfo{year}{2022}), \eprint{2211.00634}.
	
	\bibitem[{\citenamefont{Brdar and Li}(2023)}]{Vedran}
	\bibinfo{author}{\bibfnamefont{V.}~\bibnamefont{Brdar}} \bibnamefont{and}
	\bibinfo{author}{\bibfnamefont{Y.-Y.} \bibnamefont{Li}},
	\bibinfo{journal}{Phys. Lett. B} \textbf{\bibinfo{volume}{839}},
	\bibinfo{pages}{137763} (\bibinfo{year}{2023}), \eprint{2211.02028}.
	
	\bibitem[{\citenamefont{Guo et~al.}(2023)\citenamefont{Guo, Khlopov, Wu, and
			Zhu}}]{Guo}
	\bibinfo{author}{\bibfnamefont{S.-Y.} \bibnamefont{Guo}},
	\bibinfo{author}{\bibfnamefont{M.}~\bibnamefont{Khlopov}},
	\bibinfo{author}{\bibfnamefont{L.}~\bibnamefont{Wu}}, \bibnamefont{and}
	\bibinfo{author}{\bibfnamefont{B.}~\bibnamefont{Zhu}} (\bibinfo{year}{2023}),
	\eprint{2301.03523}.
	
	\bibitem[{\citenamefont{Balaji et~al.}(2023)\citenamefont{Balaji,
			Ramirez-Quezada, Silk, and Zhang}}]{Silk}
	\bibinfo{author}{\bibfnamefont{S.}~\bibnamefont{Balaji}},
	\bibinfo{author}{\bibfnamefont{M.~E.} \bibnamefont{Ramirez-Quezada}},
	\bibinfo{author}{\bibfnamefont{J.}~\bibnamefont{Silk}}, \bibnamefont{and}
	\bibinfo{author}{\bibfnamefont{Y.}~\bibnamefont{Zhang}},
	\bibinfo{journal}{Phys. Rev. D} \textbf{\bibinfo{volume}{107}},
	\bibinfo{pages}{083038} (\bibinfo{year}{2023}), \eprint{2301.02258}.
	
	\bibitem[{\citenamefont{Zhang et~al.}(2023)\citenamefont{Zhang, Murase, Ioka,
			Song, Yuan, and M\'esz\'aros}}]{Zhang:2022lff}
	\bibinfo{author}{\bibfnamefont{B.~T.} \bibnamefont{Zhang}},
	\bibinfo{author}{\bibfnamefont{K.}~\bibnamefont{Murase}},
	\bibinfo{author}{\bibfnamefont{K.}~\bibnamefont{Ioka}},
	\bibinfo{author}{\bibfnamefont{D.}~\bibnamefont{Song}},
	\bibinfo{author}{\bibfnamefont{C.}~\bibnamefont{Yuan}}, \bibnamefont{and}
	\bibinfo{author}{\bibfnamefont{P.}~\bibnamefont{M\'esz\'aros}},
	\bibinfo{journal}{Astrophys. J. Lett.} \textbf{\bibinfo{volume}{947}},
	\bibinfo{pages}{L14} (\bibinfo{year}{2023}), \eprint{2211.05754}.
	
	\bibitem[{\citenamefont{Alves~Batista}(2022)}]{AlvesBatista:2022kpg}
	\bibinfo{author}{\bibfnamefont{R.}~\bibnamefont{Alves~Batista}}
	(\bibinfo{year}{2022}), \eprint{2210.12855}.
	
	\bibitem[{\citenamefont{Gonzalez et~al.}(2023)\citenamefont{Gonzalez, Rojas,
			Pratts, Hernandez-Cadena, Fraija, Alfaro, Araujo, and
			Montes}}]{Gonzalez:2022opy}
	\bibinfo{author}{\bibfnamefont{M.~M.} \bibnamefont{Gonzalez}},
	\bibinfo{author}{\bibfnamefont{D.~A.} \bibnamefont{Rojas}},
	\bibinfo{author}{\bibfnamefont{A.}~\bibnamefont{Pratts}},
	\bibinfo{author}{\bibfnamefont{S.}~\bibnamefont{Hernandez-Cadena}},
	\bibinfo{author}{\bibfnamefont{N.}~\bibnamefont{Fraija}},
	\bibinfo{author}{\bibfnamefont{R.}~\bibnamefont{Alfaro}},
	\bibinfo{author}{\bibfnamefont{Y.~P.} \bibnamefont{Araujo}},
	\bibnamefont{and} \bibinfo{author}{\bibfnamefont{J.~A.}
		\bibnamefont{Montes}}, \bibinfo{journal}{Astrophys. J.}
	\textbf{\bibinfo{volume}{944}}, \bibinfo{pages}{178} (\bibinfo{year}{2023}),
	\eprint{2210.15857}.
	
	\bibitem[{\citenamefont{Spergel and Steinhardt}(2000)}]{Spergel:1999mh}
	\bibinfo{author}{\bibfnamefont{D.~N.} \bibnamefont{Spergel}} \bibnamefont{and}
	\bibinfo{author}{\bibfnamefont{P.~J.} \bibnamefont{Steinhardt}},
	\bibinfo{journal}{Phys. Rev. Lett.} \textbf{\bibinfo{volume}{84}},
	\bibinfo{pages}{3760} (\bibinfo{year}{2000}), \eprint{astro-ph/9909386}.
	
	\bibitem[{\citenamefont{Tulin et~al.}(2013{\natexlab{a}})\citenamefont{Tulin,
			Yu, and Zurek}}]{Tulin:2013teo}
	\bibinfo{author}{\bibfnamefont{S.}~\bibnamefont{Tulin}},
	\bibinfo{author}{\bibfnamefont{H.-B.} \bibnamefont{Yu}}, \bibnamefont{and}
	\bibinfo{author}{\bibfnamefont{K.~M.} \bibnamefont{Zurek}},
	\bibinfo{journal}{Phys. Rev. D} \textbf{\bibinfo{volume}{87}},
	\bibinfo{pages}{115007} (\bibinfo{year}{2013}{\natexlab{a}}),
	\eprint{1302.3898}.
	
	\bibitem[{\citenamefont{Kamada et~al.}(2017)\citenamefont{Kamada, Kaplinghat,
			Pace, and Yu}}]{Kamada:2016euw}
	\bibinfo{author}{\bibfnamefont{A.}~\bibnamefont{Kamada}},
	\bibinfo{author}{\bibfnamefont{M.}~\bibnamefont{Kaplinghat}},
	\bibinfo{author}{\bibfnamefont{A.~B.} \bibnamefont{Pace}}, \bibnamefont{and}
	\bibinfo{author}{\bibfnamefont{H.-B.} \bibnamefont{Yu}},
	\bibinfo{journal}{Phys. Rev. Lett.} \textbf{\bibinfo{volume}{119}},
	\bibinfo{pages}{111102} (\bibinfo{year}{2017}), \eprint{1611.02716}.
	
	\bibitem[{\citenamefont{Creasey et~al.}(2017)\citenamefont{Creasey, Sameie,
			Sales, Yu, Vogelsberger, and Zavala}}]{Creasey:2017qxc}
	\bibinfo{author}{\bibfnamefont{P.}~\bibnamefont{Creasey}},
	\bibinfo{author}{\bibfnamefont{O.}~\bibnamefont{Sameie}},
	\bibinfo{author}{\bibfnamefont{L.~V.} \bibnamefont{Sales}},
	\bibinfo{author}{\bibfnamefont{H.-B.} \bibnamefont{Yu}},
	\bibinfo{author}{\bibfnamefont{M.}~\bibnamefont{Vogelsberger}},
	\bibnamefont{and} \bibinfo{author}{\bibfnamefont{J.}~\bibnamefont{Zavala}},
	\bibinfo{journal}{Mon. Not. Roy. Astron. Soc.}
	\textbf{\bibinfo{volume}{468}}, \bibinfo{pages}{2283} (\bibinfo{year}{2017}),
	\eprint{1612.03903}.
	
	\bibitem[{\citenamefont{Ren et~al.}(2019)\citenamefont{Ren, Kwa, Kaplinghat,
			and Yu}}]{Ren:2018jpt}
	\bibinfo{author}{\bibfnamefont{T.}~\bibnamefont{Ren}},
	\bibinfo{author}{\bibfnamefont{A.}~\bibnamefont{Kwa}},
	\bibinfo{author}{\bibfnamefont{M.}~\bibnamefont{Kaplinghat}},
	\bibnamefont{and} \bibinfo{author}{\bibfnamefont{H.-B.} \bibnamefont{Yu}},
	\bibinfo{journal}{Phys. Rev. X} \textbf{\bibinfo{volume}{9}},
	\bibinfo{pages}{031020} (\bibinfo{year}{2019}), \eprint{1808.05695}.
	
	\bibitem[{\citenamefont{Agnes et~al.}(2022)}]{GlobalArgonDarkMatter:2022xgs}
	\bibinfo{author}{\bibfnamefont{P.}~\bibnamefont{Agnes}} \bibnamefont{et~al.}
	(\bibinfo{collaboration}{Global Argon Dark Matter}) (\bibinfo{year}{2022}),
	\eprint{2209.01177}.
	
	\bibitem[{\citenamefont{Aalbers et~al.}(2022)}]{LUX-ZEPLIN:2022qhg}
	\bibinfo{author}{\bibfnamefont{J.}~\bibnamefont{Aalbers}} \bibnamefont{et~al.}
	(\bibinfo{collaboration}{LUX-ZEPLIN}) (\bibinfo{year}{2022}),
	\eprint{2207.03764}.
	
	\bibitem[{\citenamefont{Borah et~al.}(2022{\natexlab{a}})\citenamefont{Borah,
			Dutta, Mahapatra, and Sahu}}]{Borah:2021yek}
	\bibinfo{author}{\bibfnamefont{D.}~\bibnamefont{Borah}},
	\bibinfo{author}{\bibfnamefont{M.}~\bibnamefont{Dutta}},
	\bibinfo{author}{\bibfnamefont{S.}~\bibnamefont{Mahapatra}},
	\bibnamefont{and} \bibinfo{author}{\bibfnamefont{N.}~\bibnamefont{Sahu}},
	\bibinfo{journal}{Nucl. Phys. B} \textbf{\bibinfo{volume}{979}},
	\bibinfo{pages}{115787} (\bibinfo{year}{2022}{\natexlab{a}}),
	\eprint{2107.13176}.
	
	\bibitem[{\citenamefont{Borah et~al.}(2021{\natexlab{a}})\citenamefont{Borah,
			Dutta, Mahapatra, and Sahu}}]{Borah:2021pet}
	\bibinfo{author}{\bibfnamefont{D.}~\bibnamefont{Borah}},
	\bibinfo{author}{\bibfnamefont{M.}~\bibnamefont{Dutta}},
	\bibinfo{author}{\bibfnamefont{S.}~\bibnamefont{Mahapatra}},
	\bibnamefont{and} \bibinfo{author}{\bibfnamefont{N.}~\bibnamefont{Sahu}}
	(\bibinfo{year}{2021}{\natexlab{a}}), \eprint{2110.00021}.
	
	\bibitem[{\citenamefont{Borah et~al.}(2021{\natexlab{b}})\citenamefont{Borah,
			Dutta, Mahapatra, and Sahu}}]{Borah:2021rbx}
	\bibinfo{author}{\bibfnamefont{D.}~\bibnamefont{Borah}},
	\bibinfo{author}{\bibfnamefont{M.}~\bibnamefont{Dutta}},
	\bibinfo{author}{\bibfnamefont{S.}~\bibnamefont{Mahapatra}},
	\bibnamefont{and} \bibinfo{author}{\bibfnamefont{N.}~\bibnamefont{Sahu}}
	(\bibinfo{year}{2021}{\natexlab{b}}), \eprint{2112.06847}.
	
	\bibitem[{\citenamefont{Borah et~al.}(2022{\natexlab{b}})\citenamefont{Borah,
			Dasgupta, Mahapatra, and Sahu}}]{Borah:2021qmi}
	\bibinfo{author}{\bibfnamefont{D.}~\bibnamefont{Borah}},
	\bibinfo{author}{\bibfnamefont{A.}~\bibnamefont{Dasgupta}},
	\bibinfo{author}{\bibfnamefont{S.}~\bibnamefont{Mahapatra}},
	\bibnamefont{and} \bibinfo{author}{\bibfnamefont{N.}~\bibnamefont{Sahu}},
	\bibinfo{journal}{Phys. Rev. D} \textbf{\bibinfo{volume}{106}},
	\bibinfo{pages}{095028} (\bibinfo{year}{2022}{\natexlab{b}}),
	\eprint{2112.14786}.
	
	\bibitem[{\citenamefont{Borah et~al.}(2022{\natexlab{c}})\citenamefont{Borah,
			Mahapatra, and Sahu}}]{Borah:2022ask}
	\bibinfo{author}{\bibfnamefont{D.}~\bibnamefont{Borah}},
	\bibinfo{author}{\bibfnamefont{S.}~\bibnamefont{Mahapatra}},
	\bibnamefont{and} \bibinfo{author}{\bibfnamefont{N.}~\bibnamefont{Sahu}}
	(\bibinfo{year}{2022}{\natexlab{c}}), \eprint{2211.15703}.
	
	\bibitem[{\citenamefont{Ibe et~al.}(2022)\citenamefont{Ibe, Kobayashi,
			Nakayama, and Shirai}}]{Ibe:2021fed}
	\bibinfo{author}{\bibfnamefont{M.}~\bibnamefont{Ibe}},
	\bibinfo{author}{\bibfnamefont{S.}~\bibnamefont{Kobayashi}},
	\bibinfo{author}{\bibfnamefont{Y.}~\bibnamefont{Nakayama}}, \bibnamefont{and}
	\bibinfo{author}{\bibfnamefont{S.}~\bibnamefont{Shirai}},
	\bibinfo{journal}{JHEP} \textbf{\bibinfo{volume}{03}}, \bibinfo{pages}{198}
	(\bibinfo{year}{2022}), \eprint{2112.11096}.
	
	\bibitem[{\citenamefont{Elor et~al.}(2023)\citenamefont{Elor, McGehee, and
			Pierce}}]{Elor:2021swj}
	\bibinfo{author}{\bibfnamefont{G.}~\bibnamefont{Elor}},
	\bibinfo{author}{\bibfnamefont{R.}~\bibnamefont{McGehee}}, \bibnamefont{and}
	\bibinfo{author}{\bibfnamefont{A.}~\bibnamefont{Pierce}},
	\bibinfo{journal}{Phys. Rev. Lett.} \textbf{\bibinfo{volume}{130}},
	\bibinfo{pages}{031803} (\bibinfo{year}{2023}), \eprint{2112.03920}.
	
	\bibitem[{\citenamefont{Cohen et~al.}(2008)\citenamefont{Cohen, Morrissey, and
			Pierce}}]{Cohen:2008nb}
	\bibinfo{author}{\bibfnamefont{T.}~\bibnamefont{Cohen}},
	\bibinfo{author}{\bibfnamefont{D.~E.} \bibnamefont{Morrissey}},
	\bibnamefont{and} \bibinfo{author}{\bibfnamefont{A.}~\bibnamefont{Pierce}},
	\bibinfo{journal}{Phys. Rev. D} \textbf{\bibinfo{volume}{78}},
	\bibinfo{pages}{111701} (\bibinfo{year}{2008}), \eprint{0808.3994}.
	
	\bibitem[{\citenamefont{Hashino et~al.}(2022)\citenamefont{Hashino, Liu, Wang,
			and Xie}}]{Hashino:2021dvx}
	\bibinfo{author}{\bibfnamefont{K.}~\bibnamefont{Hashino}},
	\bibinfo{author}{\bibfnamefont{J.}~\bibnamefont{Liu}},
	\bibinfo{author}{\bibfnamefont{X.-P.} \bibnamefont{Wang}}, \bibnamefont{and}
	\bibinfo{author}{\bibfnamefont{K.-P.} \bibnamefont{Xie}},
	\bibinfo{journal}{Phys. Rev. D} \textbf{\bibinfo{volume}{105}},
	\bibinfo{pages}{055009} (\bibinfo{year}{2022}), \eprint{2109.07479}.
	
	\bibitem[{\citenamefont{Hambye and Vanderheyden}(2020)}]{Hambye:2019tjt}
	\bibinfo{author}{\bibfnamefont{T.}~\bibnamefont{Hambye}} \bibnamefont{and}
	\bibinfo{author}{\bibfnamefont{L.}~\bibnamefont{Vanderheyden}},
	\bibinfo{journal}{JCAP} \textbf{\bibinfo{volume}{05}}, \bibinfo{pages}{001}
	(\bibinfo{year}{2020}), \eprint{1912.11708}.
	
	\bibitem[{\citenamefont{Krnjaic}(2016)}]{Krnjaic}
	\bibinfo{author}{\bibfnamefont{G.}~\bibnamefont{Krnjaic}},
	\bibinfo{journal}{Phys. Rev. D} \textbf{\bibinfo{volume}{94}},
	\bibinfo{pages}{073009} (\bibinfo{year}{2016}), \eprint{1512.04119}.
	
	\bibitem[{\citenamefont{Dev et~al.}(2022)\citenamefont{Dev, Fortin, Harris,
			Sinha, and Zhang}}]{Dev22}
	\bibinfo{author}{\bibfnamefont{P.~S.~B.} \bibnamefont{Dev}},
	\bibinfo{author}{\bibfnamefont{J.-F.} \bibnamefont{Fortin}},
	\bibinfo{author}{\bibfnamefont{S.~P.} \bibnamefont{Harris}},
	\bibinfo{author}{\bibfnamefont{K.}~\bibnamefont{Sinha}}, \bibnamefont{and}
	\bibinfo{author}{\bibfnamefont{Y.}~\bibnamefont{Zhang}},
	\bibinfo{journal}{JCAP} \textbf{\bibinfo{volume}{01}}, \bibinfo{pages}{006}
	(\bibinfo{year}{2022}), \eprint{2111.05852}.
	
	\bibitem[{\citenamefont{Dev et~al.}(2020)\citenamefont{Dev, Mohapatra, and
			Zhang}}]{Dev20}
	\bibinfo{author}{\bibfnamefont{P.~S.~B.} \bibnamefont{Dev}},
	\bibinfo{author}{\bibfnamefont{R.~N.} \bibnamefont{Mohapatra}},
	\bibnamefont{and} \bibinfo{author}{\bibfnamefont{Y.}~\bibnamefont{Zhang}},
	\bibinfo{journal}{JCAP} \textbf{\bibinfo{volume}{08}}, \bibinfo{pages}{003}
	(\bibinfo{year}{2020}), \bibinfo{note}{[Erratum: JCAP 11, E01 (2020)]},
	\eprint{2005.00490}.
	
	\bibitem[{\citenamefont{Ishizuka and Yoshimura}(1990)}]{Ishizuka}
	\bibinfo{author}{\bibfnamefont{N.}~\bibnamefont{Ishizuka}} \bibnamefont{and}
	\bibinfo{author}{\bibfnamefont{M.}~\bibnamefont{Yoshimura}},
	\bibinfo{journal}{Prog. Theor. Phys.} \textbf{\bibinfo{volume}{84}},
	\bibinfo{pages}{233} (\bibinfo{year}{1990}).
	
	\bibitem[{\citenamefont{Arndt and Fox}(2003)}]{Arndt}
	\bibinfo{author}{\bibfnamefont{D.}~\bibnamefont{Arndt}} \bibnamefont{and}
	\bibinfo{author}{\bibfnamefont{P.~J.} \bibnamefont{Fox}},
	\bibinfo{journal}{JHEP} \textbf{\bibinfo{volume}{02}}, \bibinfo{pages}{036}
	(\bibinfo{year}{2003}), \eprint{hep-ph/0207098}.
	
	\bibitem[{\citenamefont{Diener and Burgess}(2013)}]{Diener}
	\bibinfo{author}{\bibfnamefont{R.}~\bibnamefont{Diener}} \bibnamefont{and}
	\bibinfo{author}{\bibfnamefont{C.~P.} \bibnamefont{Burgess}},
	\bibinfo{journal}{JHEP} \textbf{\bibinfo{volume}{05}}, \bibinfo{pages}{078}
	(\bibinfo{year}{2013}), \eprint{1302.6486}.
	
	\bibitem[{\citenamefont{Tu and Ng}(2017)}]{Tu}
	\bibinfo{author}{\bibfnamefont{H.}~\bibnamefont{Tu}} \bibnamefont{and}
	\bibinfo{author}{\bibfnamefont{K.-W.} \bibnamefont{Ng}},
	\bibinfo{journal}{JHEP} \textbf{\bibinfo{volume}{07}}, \bibinfo{pages}{108}
	(\bibinfo{year}{2017}), \eprint{1706.08340}.
	
	\bibitem[{\citenamefont{Lee}(2018)}]{Lee}
	\bibinfo{author}{\bibfnamefont{J.~S.} \bibnamefont{Lee}}
	(\bibinfo{year}{2018}), \eprint{1808.10136}.
	
	\bibitem[{\citenamefont{Dev et~al.}(2017)\citenamefont{Dev, Mohapatra, and
			Zhang}}]{Dev17}
	\bibinfo{author}{\bibfnamefont{P.~S.~B.} \bibnamefont{Dev}},
	\bibinfo{author}{\bibfnamefont{R.~N.} \bibnamefont{Mohapatra}},
	\bibnamefont{and} \bibinfo{author}{\bibfnamefont{Y.}~\bibnamefont{Zhang}},
	\bibinfo{journal}{Nucl. Phys. B} \textbf{\bibinfo{volume}{923}},
	\bibinfo{pages}{179} (\bibinfo{year}{2017}), \eprint{1703.02471}.
	
	\bibitem[{\citenamefont{Kaplinghat et~al.}(2016)\citenamefont{Kaplinghat,
			Tulin, and Yu}}]{Kaplinghat:2015aga}
	\bibinfo{author}{\bibfnamefont{M.}~\bibnamefont{Kaplinghat}},
	\bibinfo{author}{\bibfnamefont{S.}~\bibnamefont{Tulin}}, \bibnamefont{and}
	\bibinfo{author}{\bibfnamefont{H.-B.} \bibnamefont{Yu}},
	\bibinfo{journal}{Phys. Rev. Lett.} \textbf{\bibinfo{volume}{116}},
	\bibinfo{pages}{041302} (\bibinfo{year}{2016}), \eprint{1508.03339}.
	
	\bibitem[{\citenamefont{Kamada et~al.}(2020)\citenamefont{Kamada, Kim, and
			Kuwahara}}]{Kamada:2020buc}
	\bibinfo{author}{\bibfnamefont{A.}~\bibnamefont{Kamada}},
	\bibinfo{author}{\bibfnamefont{H.~J.} \bibnamefont{Kim}}, \bibnamefont{and}
	\bibinfo{author}{\bibfnamefont{T.}~\bibnamefont{Kuwahara}},
	\bibinfo{journal}{JHEP} \textbf{\bibinfo{volume}{12}}, \bibinfo{pages}{202}
	(\bibinfo{year}{2020}), \eprint{2007.15522}.
	
	\bibitem[{\citenamefont{Tulin and Yu}(2018)}]{Tulin:2017ara}
	\bibinfo{author}{\bibfnamefont{S.}~\bibnamefont{Tulin}} \bibnamefont{and}
	\bibinfo{author}{\bibfnamefont{H.-B.} \bibnamefont{Yu}},
	\bibinfo{journal}{Phys. Rept.} \textbf{\bibinfo{volume}{730}},
	\bibinfo{pages}{1} (\bibinfo{year}{2018}), \eprint{1705.02358}.
	
	\bibitem[{\citenamefont{Tulin et~al.}(2013{\natexlab{b}})\citenamefont{Tulin,
			Yu, and Zurek}}]{Tulin:2012wi}
	\bibinfo{author}{\bibfnamefont{S.}~\bibnamefont{Tulin}},
	\bibinfo{author}{\bibfnamefont{H.-B.} \bibnamefont{Yu}}, \bibnamefont{and}
	\bibinfo{author}{\bibfnamefont{K.~M.} \bibnamefont{Zurek}},
	\bibinfo{journal}{Phys. Rev. Lett.} \textbf{\bibinfo{volume}{110}},
	\bibinfo{pages}{111301} (\bibinfo{year}{2013}{\natexlab{b}}),
	\eprint{1210.0900}.
	
	\bibitem[{\citenamefont{Khrapak
			et~al.}(2003{\natexlab{a}})\citenamefont{Khrapak, Ivlev, Morfill, and
			Zhdanov}}]{Khrapak:2003kjw}
	\bibinfo{author}{\bibfnamefont{S.~A.} \bibnamefont{Khrapak}},
	\bibinfo{author}{\bibfnamefont{A.~V.} \bibnamefont{Ivlev}},
	\bibinfo{author}{\bibfnamefont{G.~E.} \bibnamefont{Morfill}},
	\bibnamefont{and} \bibinfo{author}{\bibfnamefont{S.~K.}
		\bibnamefont{Zhdanov}}, \bibinfo{journal}{Phys. Rev. Lett.}
	\textbf{\bibinfo{volume}{90}}, \bibinfo{pages}{225002}
	(\bibinfo{year}{2003}{\natexlab{a}}).
	
	\bibitem[{\citenamefont{Borah et~al.}(2022{\natexlab{d}})\citenamefont{Borah,
			Dutta, Mahapatra, and Sahu}}]{BorahD}
	\bibinfo{author}{\bibfnamefont{D.}~\bibnamefont{Borah}},
	\bibinfo{author}{\bibfnamefont{M.}~\bibnamefont{Dutta}},
	\bibinfo{author}{\bibfnamefont{S.}~\bibnamefont{Mahapatra}},
	\bibnamefont{and} \bibinfo{author}{\bibfnamefont{N.}~\bibnamefont{Sahu}},
	\bibinfo{journal}{Phys. Rev. D} \textbf{\bibinfo{volume}{105}},
	\bibinfo{pages}{075019} (\bibinfo{year}{2022}{\natexlab{d}}),
	\eprint{2112.06847}.
	
	\bibitem[{\citenamefont{Abdelhameed et~al.}(2019)}]{CRESST:2019jnq}
	\bibinfo{author}{\bibfnamefont{A.~H.} \bibnamefont{Abdelhameed}}
	\bibnamefont{et~al.} (\bibinfo{collaboration}{CRESST}),
	\bibinfo{journal}{Phys. Rev. D} \textbf{\bibinfo{volume}{100}},
	\bibinfo{pages}{102002} (\bibinfo{year}{2019}), \eprint{1904.00498}.
	
	\bibitem[{\citenamefont{Billard et~al.}(2022)}]{Billard:2021uyg}
	\bibinfo{author}{\bibfnamefont{J.}~\bibnamefont{Billard}} \bibnamefont{et~al.},
	\bibinfo{journal}{Rept. Prog. Phys.} \textbf{\bibinfo{volume}{85}},
	\bibinfo{pages}{056201} (\bibinfo{year}{2022}), \eprint{2104.07634}.
	
	\bibitem[{\citenamefont{Ibe et~al.}(2018)\citenamefont{Ibe, Nakano, Shoji, and
			Suzuki}}]{Ibe:2017yqa}
	\bibinfo{author}{\bibfnamefont{M.}~\bibnamefont{Ibe}},
	\bibinfo{author}{\bibfnamefont{W.}~\bibnamefont{Nakano}},
	\bibinfo{author}{\bibfnamefont{Y.}~\bibnamefont{Shoji}}, \bibnamefont{and}
	\bibinfo{author}{\bibfnamefont{K.}~\bibnamefont{Suzuki}},
	\bibinfo{journal}{JHEP} \textbf{\bibinfo{volume}{03}}, \bibinfo{pages}{194}
	(\bibinfo{year}{2018}), \eprint{1707.07258}.
	
	\bibitem[{\citenamefont{Arnaud et~al.}(2018)}]{NEWS-G:2017pxg}
	\bibinfo{author}{\bibfnamefont{Q.}~\bibnamefont{Arnaud}} \bibnamefont{et~al.}
	(\bibinfo{collaboration}{NEWS-G}), \bibinfo{journal}{Astropart. Phys.}
	\textbf{\bibinfo{volume}{97}}, \bibinfo{pages}{54} (\bibinfo{year}{2018}),
	\eprint{1706.04934}.
	
	\bibitem[{\citenamefont{Agnese et~al.}(2017)}]{SuperCDMS:2016wui}
	\bibinfo{author}{\bibfnamefont{R.}~\bibnamefont{Agnese}} \bibnamefont{et~al.}
	(\bibinfo{collaboration}{SuperCDMS}), \bibinfo{journal}{Phys. Rev. D}
	\textbf{\bibinfo{volume}{95}}, \bibinfo{pages}{082002}
	(\bibinfo{year}{2017}), \eprint{1610.00006}.
	
	\bibitem[{\citenamefont{Agnes et~al.}(2023)}]{GlobalArgonDarkMatter:2022ppc}
	\bibinfo{author}{\bibfnamefont{P.}~\bibnamefont{Agnes}} \bibnamefont{et~al.}
	(\bibinfo{collaboration}{Global Argon Dark Matter}), \bibinfo{journal}{Phys.
		Rev. D} \textbf{\bibinfo{volume}{107}}, \bibinfo{pages}{112006}
	(\bibinfo{year}{2023}), \eprint{2209.01177}.
	
	\bibitem[{\citenamefont{Balaji et~al.}(2022)\citenamefont{Balaji, Dev, Silk,
			and Zhang}}]{Balaji22}
	\bibinfo{author}{\bibfnamefont{S.}~\bibnamefont{Balaji}},
	\bibinfo{author}{\bibfnamefont{P.~S.~B.} \bibnamefont{Dev}},
	\bibinfo{author}{\bibfnamefont{J.}~\bibnamefont{Silk}}, \bibnamefont{and}
	\bibinfo{author}{\bibfnamefont{Y.}~\bibnamefont{Zhang}},
	\bibinfo{journal}{JCAP} \textbf{\bibinfo{volume}{12}}, \bibinfo{pages}{024}
	(\bibinfo{year}{2022}), \eprint{2205.01669}.
	
	\bibitem[{\citenamefont{Khrapak
			et~al.}(2003{\natexlab{b}})\citenamefont{Khrapak, Ivlev, Morfill, and
			Zhdanov}}]{PhysRevLett.90.225002}
	\bibinfo{author}{\bibfnamefont{S.~A.} \bibnamefont{Khrapak}},
	\bibinfo{author}{\bibfnamefont{A.~V.} \bibnamefont{Ivlev}},
	\bibinfo{author}{\bibfnamefont{G.~E.} \bibnamefont{Morfill}},
	\bibnamefont{and} \bibinfo{author}{\bibfnamefont{S.~K.}
		\bibnamefont{Zhdanov}}, \bibinfo{journal}{Phys. Rev. Lett.}
	\textbf{\bibinfo{volume}{90}}, \bibinfo{pages}{225002}
	(\bibinfo{year}{2003}{\natexlab{b}}),
	
\end{thebibliography}

\end{document}